\documentclass[sigplan,screen]{acmart}

\setcopyright{cc}
\setcctype{by}
\acmDOI{10.1145/3735950.3735959}
\acmYear{2025}
\copyrightyear{2025}
\acmISBN{979-8-4007-1610-2/25/06}
\acmConference[ISMM '25]{Proceedings of the 2025 ACM SIGPLAN International Symposium on Memory Management}{June 17, 2025}{Seoul, Republic of Korea}
\acmBooktitle{Proceedings of the 2025 ACM SIGPLAN International Symposium on Memory Management (ISMM '25), June 17, 2025, Seoul, Republic of Korea}
\received{2025-03-19}
\received[accepted]{2025-05-03}

\usepackage[english]{babel}
\usepackage{fancyvrb}
\usepackage{enumitem}
\usepackage{xspace}
\usepackage{tikz}
\usetikzlibrary{patterns}
\usepackage{pifont}
\usepackage{multirow}
\usepackage{tcolorbox}
\usepackage{bm}
\usepackage{pdfpages}
\usepackage{microtype}
\usepackage{pgfplots}
\usepackage{tikz}
\usepackage{amsthm}
\usepackage[normalem]{ulem}
\usepackage{wrapfig}
\usepackage{svg}
\usepackage{booktabs}
\usepackage{siunitx,array,threeparttable}
\usepackage{comment}
\usepackage{balance}
\usepackage{microtype}

\DeclareMathAlphabet{\mathcal}{OMS}{cmsy}{m}{n}

\newcommand{\ignore}[1]{}%
\newcommand{\myparagraph}[1]{\vspace{0.5em}\noindent\emph{#1}.}
\newcommand{\decodePtr}[0]{\mathit{decode}\xspace}

\newcommand{\isEncoded}[0]{\mathit{isEncoded}\xspace}
\newcommand{\assignId}[0]{\mathit{generateId}\xspace}
\newcommand{\mapping}[0]{\mathcal{M}\xspace}

\newcommand{\tool}[0]{\textsc{BlueFat}\xspace}
\newcommand{\bluefat}[0]{\tool}
\newcommand{\greenfat}[0]{\textsc{GreenFat}\xspace}
\newcommand{\FRP}[0]{FRP\xspace}
\newcommand*\rot{\rotatebox{90}}
\newcommand*{\p}[1]{\begin{tikzpicture}[scale=0.08]%
    \draw (0,0) circle (1);
    \fill[fill=black] (0,0) -- (90:1) arc (90:90-#1*3.6:1) -- cycle;
    \end{tikzpicture}}
\newcommand{\cmark}[0]{\color{green!50!black}{\ding{51}}}
\newcommand{\xmark}[0]{\color{red!50!black}{\ding{55}}}

\theoremstyle{definition}
\newtheorem{example}{Example}[]

\usepackage{amsmath}
\usepackage{graphicx}

\title{Fully Randomized Pointers}

\begin{abstract}
Memory errors continue to be a critical concern for programs written in low-level programming languages such as \texttt{C} and \texttt{C++}.
Many different memory error defenses have been proposed, each with varying trade-offs in terms of overhead, compatibility, and attack resistance.
Some defenses are highly compatible but only provide minimal protection, and can be easily bypassed by knowledgeable attackers.
On the other end of the spectrum, capability systems offer very strong (unforgeable) protection, but require novel software and hardware implementations that are incompatible by definition.
The challenge is to achieve {\em both} very strong protection and high compatibility.

In this paper, we propose {\em Fully Randomized Pointers} (\FRP{}) as a strong memory error defense that also maintains compatibility with existing binary software.
The key idea behind \FRP{} is to design a new pointer encoding scheme that allows for the full randomization of most pointer bits, rendering even brute force attacks impractical. 
We design a \FRP{} encoding that is: (1) compatible with existing binary code (recompilation not needed);
and (2) decoupled from the underlying object layout.
\FRP{} is prototyped as: 
(i) a software implementation (\tool) to test security and compatibility; and 
(ii) a proof-of-concept hardware implementation (\greenfat) to evaluate performance.
We show \FRP{} is secure, practical, and compatible at the binary level, while our hardware implementation achieves low performance overheads (${<}4\%$).
\end{abstract}

\begin{document}
\date{}

\author{Sai Dhawal Phaye}
\orcid{0009-0005-5960-0574}
\affiliation{%
  \institution{National University of Singapore}
   \country{Singapore}
}
\email{sdp@comp.nus.edu.sg}

\author{Gregory J. Duck}
\orcid{0000-0002-0837-9671}
\authornote{Joint first author}
\affiliation{%
  \institution{National University of Singapore}
   \country{Singapore}
}
\email{gregory@comp.nus.edu.sg}

\author{Roland H. C. Yap}
\orcid{0000-0002-1188-7474}
\affiliation{%
  \institution{National University of Singapore}
   \country{Singapore}
}
\email{ryap@comp.nus.edu.sg}

\author{Trevor E. Carlson}
\orcid{0000-0001-8742-134X}
\affiliation{%
  \institution{National University of Singapore}
  \country{Singapore}
}
\email{tcarlson@comp.nus.edu.sg}

\begin{CCSXML}
<ccs2012>
<concept>
<concept_id>10002978.10003022.10003023</concept_id>
<concept_desc>Security and privacy~Software security engineering</concept_desc>
<concept_significance>500</concept_significance>
</concept>
<concept>
<concept_id>10002978.10003001.10003599</concept_id>
<concept_desc>Security and privacy~Hardware security implementation</concept_desc>
<concept_significance>500</concept_significance>
</concept>
</ccs2012>
\end{CCSXML}

\ccsdesc[500]{Security and privacy~Software security engineering}
\ccsdesc[500]{Security and privacy~Hardware security implementation}

\keywords{Memory safety; fully randomized pointers}

\maketitle

\section{Introduction}

Memory errors, such as buffer overflows and use-after-free bugs, 
still continue to be the primary cause of security vulnerabilities in
software implemented in low-level programming languages such as \verb+C+ and \verb_C++_~\cite{chromium-memory,microsoft-bluehat-talk}.
Many different tools and techniques have been proposed to defend against memory errors both in software and hardware~\cite{asan, song19sok, duck16heap, duck18effective, duck17stack, nethercote2007valgrind, bruening2011drmem, berge06diehard, gene10dieharder, duck2022redfat, lemay2021c3, nagarakatte09softbound, nagarakatte2010cets, saileshwar2022heapcheck, li2022pacmem, cheri,yu2023capstone}.
Existing state-of-the-art tools have inherent trade-offs, especially concerning performance, security, and binary compatibility.
Some tools (e.g.,~\cite{asan, duck18effective, nethercote2007valgrind, bruening2011drmem}) are primarily designed for {\em software testing} rather than security.
Others (e.g.~\cite{duck2022redfat, berge06diehard, gene10dieharder}) are primarily designed for {\em security hardening}---i.e., to prevent a proactive attacker from exploiting memory errors.

For program hardening under adversarial attacker models, the key
trade-off is {\em bypass resistance}---i.e., how easily a proactive attacker (who is aware that the program is hardened) can adapt their attack to bypass the defense?
Many existing tools have poor resistance.
For example, bug detection tools based on {\em memory poisoning}, such as AddressSanitizer (ASAN)~\cite{asan} and Valgrind~\cite{nethercote2007valgrind}, famously do not track pointer origins (a.k.a., {\em pointer provenance}~\cite{memarian2019ptr}).
While this does achieve good compatibility, it nevertheless allows attackers to construct seemingly valid pointers to out-of-bounds objects~\cite{duck2022redfat}---effectively bypassing the defense.

Recently, several entropy-based memory safety solutions, based on pointer {\em tagging}, {\em authentication}, or {\em encryption}, have been proposed---offering improved bypass resistance.
One common idea is to attach a {\em Pointer Authentication Code} (PAC, e.g., PACMem~\cite{li2022pacmem}) or {\em tag}~(e.g., HeapCheck~\cite{saileshwar2022heapcheck}) to each pointer.
The tag typically resides in the upper pointer bits.
Another related idea is ($C^3$)~\cite{lemay2021c3}, which partially {\em encrypts} an encoded pointer value, rather than using separate authentication codes/tags.
The encryption hardens against tampering of the relationship between pointers and the underlying object.
Being based on hardware extensions, these solutions tend to achieve good binary compatibility and low performance overheads.
For security, we can classify these solutions as {\em entropy}-based defenses, since an attacker must determine a (randomized) tag or encrypted value in order to successfully bypass the defense.
Assuming the defense is well designed, the attacker will have no better method other than to guess the correct value, and an incorrect guess will lead to immediate program termination.
Each of PACMem, HeapCheck, and $C^3$ use \textbf{24 bits} of entropy.

Entropy-based defenses rely on the difficulty for an attacker to
guess the correct value.
However, practical attacks are still feasible when the entropy is low (e.g., 24 bits) and the key assumption (that an attacker can only make a single incorrect guess) does not hold---allowing the attacker to ``brute force'' the required value.
For example, it is not uncommon for programs to be configured to automatically restart after a crash, e.g., a network server may automatically restart in order to minimize service disruption.
Alternatively, there may be multiple vulnerable servers on a given network.
Such configurations effectively grant the attacker multiple attempts~\cite{bittau14hacking}.
Another attack vector is {\em side channels}, such as {\em speculative execution}~\cite{kocher2019spectre}, which allow the attacker to test values without causing program termination.
This approach has been proven effective against ARM pointer authentication (PAC)
~\cite{ravi2022pacman}.
Finally, well-known exploitation techniques (e.g., heap spray) can further weaken the defense.

\begin{figure}[tb]
    \centering
    \includegraphics[scale=3]{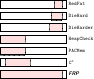}
    \caption{Entropy for each defense.  More entropy = better.\label{fig:entropy}}
\end{figure}

In this paper, we propose a new (hardware-based) memory error defense that is designed to {\em maximize} entropy, hence maximizing bypass resistance.
Our approach is based on a simple observation that most (binary) software is agnostic to the underlying pointer encoding---provided that core pointer operations (arithmetic, difference, comparison, dereference, etc.) implement the expected semantics.
We design a new encoding that represents pointers as {\em cryptographically secure random numbers}, ensuring that two pointers to different objects will have:
(i) bit patterns with no predictable relation to each other;
and (ii) no predictable relation to the underlying address(es) in memory.
For existing \verb+x86_64+ binary code, we show a baseline entropy of \textbf{52 bits} can be supported under conservative assumptions, offering robust memory error protection with significantly stronger resistance to bypass attacks (${\sim}2^{51}$ attack attempts required versus ${\sim}8$M).
The high degree of randomization provides a much stronger resistance against {\em forgeability}, i.e., the ability to construct invalid pointer values that fool the defense. 
Under relaxed assumptions, the effective entropy is further improved up to the full \textbf{64 bits}.
Our approach is illustrated in Figure~\ref{fig:entropy}.
Here, each colored inner box represents the position and scale of the bits that the attacker must guess to bypass the corresponding defense.
Our example models 60 bits of randomization, offering very strong resistance against brute force-style attacks.

We prototype a \FRP{} implementation for heap objects, including a software implementation (\bluefat) to test binary compatibility and security, and a hardware simulation (\greenfat) to test performance.
We show that \FRP{} offers very strong resistance to memory errors for {\em dynamically allocated memory} (\verb+malloc+/\verb+free+), including object bounds errors and use-after-free.
We also show that, despite using a radically different pointer encoding, \FRP{} is compatible with existing \verb+x86_64+ binary code {\em without the need for recompilation}.
In fact, the binary software is unaware that a memory error defense has even been applied, i.e., our design is fully transparent.
We also present (\greenfat) as a proof-of-concept hardware simulation of \FRP{}, achieving an overhead (${<}4\%$)---within an acceptable range for many use cases.
In summary, the main contributions of this paper are:
\begin{itemize}[leftmargin=*,noitemsep]
    \item We propose a new heap pointer encoding that is (1) decoupled from the machine addresses, and (2) fully randomized.
    The high degree of randomization hardens against memory errors and brute-force style bypass attacks, providing capability-like protection under a probabilistic guarantee.
    \item We show that the 
\FRP design is both efficient and compatible with existing \verb+x86_64+ binary code.
    \item We have implemented \FRP{} in software (\bluefat) and a hardware simulation (\greenfat).
    We show \FRP{} offers strong memory safety (spatial and temporal),
    bypass resistance against strong attackers, and high compatibility.
\end{itemize}

\subsection*{Open Source Release}

The software implementation (\tool) is available at:
\begin{itemize}
\item \url{https://github.com/GJDuck/BlueFat}
\end{itemize}

\section{Background}

In this section, we briefly summarize the threat model, existing mitigations, the problem statement, and our design.

\subsection{Memory Errors}
Low-level programming languages are vulnerable to {\em memory errors} such as {\em object bounds overflows} and {\em use-after-free}.
Under the \verb+C+/\verb_C++_ standards~\cite{iso2018cstd}, memory errors are {\em undefined behavior}, meaning that there is no prescribed semantics.
For example, consider the allocation:
{
\setlength{\abovedisplayskip}{0.3em}
\setlength{\belowdisplayskip}{0.3em}
\begin{align}
        \texttt{p = (T *)malloc(N * sizeof(T));} \label{eq:alloc} \tag{\textsc{Alloc}}
\end{align}
}%
If non-\verb+NULL+, \verb+p+ will point to the base of a new object of type \verb+T[N]+.
Under the \verb+C+ standard, the object can only be accessed via \verb+p+, or a pointer \emph{derived} from \verb+p+ using pointer arithmetic $\mathtt{p}{+}k$ for $k \in 0..N{-}1$.
For other values of $k$, access is expressly undefined (\cite{iso2018cstd} \S 6.5.6 \P 8).

Modern compilers optimize programs under the assumption that undefined behavior does not occur.
If (or when) this assumption is broken, a memory error may occur which results in an access of 
a memory location outside the intended object, possibly including memory belonging to other objects.
This behavior can be exploited.
For example, an out-of-bounds write can be used to modify a function pointer stored in another object, leading to a {\em control-flow hijacking} attack.
Similarly, an out-of-bounds read can leak sensitive information (i.e., an {\em information disclosure attack}), such as the infamous HeartBleed bug~\cite{heartbleed}.
Use-after-free errors can be similarly exploited, by accessing a dangling pointer to a \verb+free+'ed object, where the underlying memory has been reallocated.
The importance of hardening against use-after-free should not be underestimated---around 50\% of the serious vulnerabilities in Chrome are due to use-after-free \cite{chromium-memory}.

\begin{table}[tb]
\setlength{\tabcolsep}{2.4pt}
\centering
\scriptsize
\caption{Summary of memory error defenses. \label{tab:sanitizers}}
\begin{tabular}{|l|ccc|ccc|cc|r|}
\cline{2-10}
\multicolumn{1}{c|}{} &
                   \multicolumn{3}{c|}{\emph{Methodology}} &
                   \multicolumn{5}{c|}{\emph{Error Detection}} &
                   \multicolumn{1}{c|}{\emph{Entropy}} \\
\cline{2-10}
\multicolumn{1}{c|}{\emph{Defense}}
                 & \multicolumn{1}{c|}{{\em technology}}
                 & \rot{\em binary?}
                 & \rot{\em bypass resist.}                 
                 & \rot{\em overflow}
                 & \rot{\em underflow}
                 & \rot{\em use-after-free~~}
                 & \rot{\em false positives}
                 & \rot{\em false negatives~}
                 & \\
\cline{2-10}
\hline
\emph{ASLR} &
    \multicolumn{1}{c|}{\emph{randomization}} & b & \p{0} & \p{0} & \p{0} & \p{0} & - & \xmark & 
    n.a. \\
\hline
\emph{efence}~\cite{bruce1993electri} &
    \multicolumn{1}{c|}{\emph{page permissions}} & b & \p{0} & \p{75} & \p{25} & \p{100} & - & \xmark &
    n.a. \\
Valgrind~\cite{nethercote2007valgrind} & 
    \multicolumn{1}{c|}{\emph{mem. poisoning}} & b & \p{0} & \p{100} & \p{100} & \p{100} & - & \xmark &
    n.a. \\
DrMemory~\cite{bruening2011drmem} &
    \multicolumn{1}{c|}{\emph{mem. poisoning}} & b & \p{0} & \p{100} & \p{100} & \p{100} & - & \xmark &
    n.a. \\
ASAN~\cite{asan} &
    \multicolumn{1}{c|}{\emph{mem. poisoning}} & - & \p{0} & \p{100} & \p{100} & \p{100} & - & \xmark &
    n.a. \\
LowFat~\cite{duck16heap, duck17stack} & 
    \multicolumn{1}{c|}{\emph{low fat ptrs}} & - & \p{25} & \p{75} & \p{100} & \p{0} & \xmark & \xmark &
    n.a. \\
EffectiveSan~\cite{duck18effective} & 
    \multicolumn{1}{c|}{\emph{low fat ptrs}} & - & \p{25} & \p{100} & \p{100} & \p{100} & \xmark & \xmark &
    n.a. \\
SoftBound+CETS~\cite{nagarakatte09softbound, nagarakatte2010cets} &
    \multicolumn{1}{c|}{\emph{ptr metadata}} & - & \p{100} & \p{100} & \p{100} & \p{100} & - & - & n.a. \\
\hline
\textsc{RedFat}~\cite{duck2022redfat} &
    \multicolumn{1}{c|}{\emph{low fat ptrs}} & b & \p{25} & \p{100} & \p{100} & \p{100} & - & \xmark &
    {214.5} (*) \\
DieHard~\cite{berge06diehard} &
    \multicolumn{1}{c|}{\emph{randomization}} & b & \p{25} & \p{25} & \p{25} & \p{25} & - & \xmark &
    {2744.9} (*) \\
DieHarder~\cite{gene10dieharder} &
    \multicolumn{1}{c|}{\emph{randomization}} & b & \p{25} & \p{25} & \p{25} & \p{25} & - & \xmark &
    {6207.4} (*) \\
\hline
PACMem~\cite{li2022pacmem} &
    \multicolumn{1}{c|}{\emph{authentication}} & - & \p{25} & \p{100} & \p{100} & \p{100} & - & - &
    $2^{24}$ ($\dagger$) \\
HeapCheck~\cite{saileshwar2022heapcheck} &
    \multicolumn{1}{c|}{\emph{tagged ptrs}} & b & \p{25} & \p{100} & \p{100} & \p{100} & - & - &
    $2^{24}$ ($\dagger$) \\
$C^3$~\cite{lemay2021c3} &
    \multicolumn{1}{c|}{\emph{ptr encryption}} & b & \p{25} & \p{100} & \p{100} & \p{100} & - & - &
    $2^{24}$ ($\dagger$) \\
\hline
{\bf \FRP{}} &
    \multicolumn{1}{c|}{{\bf \emph{randomization}}} & {\bf b} & \p{100} & \p{100} & \p{100} & \p{100} & {\bf -} & {\bf -} &
    $\bm{2^{52}}$ ($\dagger$) \\
\hline
\multicolumn{1}{l}{} \\
\end{tabular}
~~~~
\setlength{\tabcolsep}{5pt}
\begin{tabular}{llll}
Key: &
 \emph{Methodology} & \emph{Error Detection}  & \emph{Entropy} \\
& $\p{0} = \text{none,minimal}$ & $\p{0} = \text{no support}$ & $1.0 = \text{attack attempt}$ \\
& $\p{25} = \text{weak}$ & $\p{25} = \text{partial}$ & $* = \text{measured}$  \\
& $\p{100} = \text{strong}$ & $\p{75} = \text{byte imprecise}$ & $\dagger = \text{theoretical}$ \\
& b = no source & $\p{100} = \text{full/precise}$ & n.a. = not entropy defense \\
& ~~~~~~\text{code needed} & - $=$ \text{none or} \\
& & ~~~~~~\text{highly improbable} \\
& & {\xmark} $=$ \text{occurs} \\
\end{tabular}
\vspace*{-2mm}
\end{table}

\newcommand{\capAny}[0]{\textls[-20]{\textsf{Atk.*}}\xspace}
\newcommand{\capInvalid}[0]{\textls[-20]{\textsf{Atk.Invalid}}\xspace}
\newcommand{\capDeref} [0] {\textls[-20]{\textsf{Atk.Deref}}\xspace}
\newcommand{\capLayout} [0]{\textls[-20]{\textsf{Atk.Layout}}\xspace}
\newcommand{\capRetry} [0] {\textls[-20]{\textsf{Atk.Retry}}\xspace}

\subsection{Threat Model}\label{sec:attacker}
We model a strong-yet-realistic attacker with the following core set of {\em attacker abilities} (\capAny):
\begin{enumerate}[leftmargin=2cm,noitemsep]
\item[\hfill\textbf{\capInvalid})]
       able to construct an {\em out-of-bounds} pointer $\mathtt{q}{=}\mathtt{p}{+}k$
       or retain a {\em dangling} pointer;
\item[\hfill\textbf{\capDeref})]
       able to dereference the invalid pointer \verb+q+;
\item[\hfill\textbf{\capLayout})]
       is knowledgeable of allocated objects and their relative layout in memory; and
\item[\hfill\textbf{\capRetry})]
       able to retry or launch multiple attacks, up to some reasonable resource bound.
\end{enumerate}
The attacker uses a combination of (\capAny) to engineer a memory error that {\em accesses an object of choice}.
In the case of the HeartBleed bug~\cite{heartbleed},
the attacker induces the program to {\em construct} (\capInvalid) a (bad) out-of-bounds pointer of the form ($\mathtt{buf}{+}k$) for a given buffer \verb+buf+ and value $k$ beyond the length of \verb+buf+.
Note that when the attacker can control $k$, it is easy to attack defenses with low bypass resistance.
The invalid pointer is then read-from ({\em dereferenced},~\capDeref), and the read value is copied into a reply message that is sent back to the attacker.
The attacker also exploits basic knowledge about the {\em layout} of objects in memory~(\capLayout).
For example, heap memory is {\em contiguous}, meaning that other objects---including possibly sensitive information---are stored beyond the end of the buffer.
Even if one attempt fails, the attacker may have the ability to {\em retry} attacks (\capRetry).
Some examples include:
\begin{enumerate}[leftmargin=*,noitemsep]
    \item {\em speculative execution} attacks, like PACMan~\cite{ravi2022pacman}, allow the attacker to try attacks without risking termination;
    \item a vulnerable server may restart after a crash~\cite{bittau14hacking};
    \item multiple vulnerable servers on the network;
    \item the attacker may use common exploitation techniques, such as {\em heap spraying}, to increase the number of targets.
\end{enumerate}
The ability to retry attacks (\capRetry) is very powerful, as it can significantly improve the probability of eventual success.

Both (\capInvalid), (\capDeref), and some combination of (\capLayout) or (\capRetry) are necessary for a successful attack.
For example, if the attacker were unable to construct and dereference an invalid/bad pointer ($\mathtt{buf}{+}k$), then no attack would be possible.
Similarly, (\capLayout) is also an important requirement,
since the attacker must determine the correct $k$ necessary for a successful attack.
Sometimes the attacker may only have {\em partial} knowledge, e.g., that sensitive data is ``likely'' to be beyond the end of a buffer.
However, assuming that attacks can be repeated (\capRetry), even partial knowledge may be sufficient.
We also assume that the attacker is knowledgeable about any memory error defense used by the program.
The attacker may also attempt to {\em bypass} the defense within (\capAny), and we refer to this as a {\em bypass attack}.

\subsection{Threat Mitigation}
Since memory errors are a well-known problem, many existing defenses and mitigation strategies have been proposed.
Some prominent examples are shown in Table~\ref{tab:sanitizers}.
Here, the table summarizes the claimed (from paper)
degree of error detection, the underlying error detection methodology, and (where applicable) micro-benchmarks measuring attack resistance w.r.t. our threat model.
Most existing defenses use (some combination of) {\em page-level protections}, {\em memory poisoning}, {\em low fat pointers}, {\em pointer metadata}, {\em authentication}, {\em encryption}, or {\em randomization} as the underlying methodology.
We evaluate each tool under the default configuration from their publicly available repository.

Many Table~\ref{tab:sanitizers} defenses target invalid pointer {\em construction} (\capInvalid) or {\em dereference} (\capDeref), having well-known limitations regarding compatibility and security.
For example, LowFat~\cite{duck16heap} protects pointer arithmetic (\capInvalid),
but may also falsely flag ``benign'' out-of-bounds pointers
(see {\em false positives} in Table~\ref{tab:sanitizers}).
This can result in lesser compatibility with real-world code.
Other defenses, such as Valgrind~\cite{nethercote2007valgrind}, DrMemory~\cite{bruening2011drmem}, and ASAN~\cite{asan}, target dereference (\capDeref) using {\em memory poisoning}, e.g., {\em poisoned redzones} inserted between objects.
However, these defenses do not track pointer validity, meaning that {\em valid} and {\em invalid} access to non-poisoned memory cannot be distinguished---allowing bypass attacks (see {\em false negatives} in Table~\ref{tab:sanitizers}).
Here, a pointer is {\em valid} if it points to the allocated object from which it was derived, as defined by the rules of \emph{pointer provenance}~\cite{memarian2019ptr}.
SoftBound+CETS~\cite{nagarakatte09softbound, nagarakatte2010cets} 
implements a strong defense by explicitly attaching metadata to each pointer and targets (\capDeref).
However, pointer metadata is heavyweight, not binary compatible, and does not handle some cases like integer-to-pointer casts (e.g., XOR-linked lists).

A few defenses target (\capLayout), and some do so only weakly.
{\em Address Space Layout Randomization} (ASLR) randomizes inter-region relationships, but is deterministic for intra-region. 
DieHard~\cite{berge06diehard} and DieHarder~\cite{gene10dieharder} also randomize object locations {\em within} the heap.
However, the attacker can still infer partial information (e.g., likely address range of allocated objects), so the defense is weak, especially when attacks can be retried (\capRetry).

\subsection{Problem Statement}\label{sec:problem}

Although some existing systems randomize pointers, they do so only weakly, and 
insufficiently under our threat model.
If pointers are represented as machine addresses, 
then pointer randomization necessitates layout randomization.
However, 
layout randomization is significantly limited in practice:
\begin{enumerate}[leftmargin=*,noitemsep,label=(\roman*)]
\item \label{limit:x64} The \verb+x86_64+ implements a 48 bit virtual address space, meaning that the upper 16 bits are zeroed in user-mode;
\item \label{limit:layout} Due to performance and memory constraints, objects are usually allocated (near) contiguously, i.e., many objects are ``close'' and often share the same page(s).
\end{enumerate}
To illustrate the problem, we design two proof-of-concept attacks that are allowable under our threat model: {\em overflow} $(\mathit{OF})$ and {\em underflow} $(\mathit{UF})$:

\vspace{0.2em}
{\small
\begin{Verbatim}[commandchars=\\\{\},codes={\catcode`$=3\catcode`^=7} ]
 \textbf{for} (size_t i = 0; i < N; i++) \textbf{attack}(p+1+i); $(OF)$
 \textbf{for} (size_t i = 0; i < N; i++) \textbf{attack}(p-1-i); $(UF)$
\end{Verbatim}
}
\vspace{0.2em}

\noindent
Here, \verb+p+ is a valid pointer to a word-sized object, and the \verb+attack()+ function attempts to access the given {\em invalid} (out-of-bounds) pointer to attack another object pointed to by \verb+q+. 
Under our threat model, the \verb+attack()+ function 
can effectively ignore crashes/errors, allowing for the attack to be retried (\capRetry).
The results are shown in Table~\ref{tab:sanitizers} under the ({\em Entropy}) column, representing attack resistance.
We either measure the average\footnote{Over 10,000 runs with a (worst-case) ${\pm}2.2\%$ margin for 95\% confidence.}
number of attempts necessary for a successful targeted attack (*), or determine the necessary value from theory ($\dagger$).\footnote{We state the theoretical entropy for systems where practical experimentation is not possible.}
A higher value is better.
The results show that, apart from this proposal, all of the entropy-based Table~\ref{tab:sanitizers} defenses are vulnerable under our threat model.
For example, due to limitations \ref{limit:x64} and \ref{limit:layout}, tools based on layout randomization have a best-case ${\sim}$\textbf{12.6 bits} of effective security (DieHarder).
Existing pointer authentication, tagging, and encryption schemes use \textbf{24 bits} of entropy.
Although an improvement, even this is still not sufficient against practical bypass attacks~\cite{ravi2022pacman}.
The challenge is to design a memory error defense that can resist even very strong threat models.

\subsection{Our Approach}

Memory errors are exploitable when the relationship between pointers is predictable.
Some existing defenses attempt to counter this using randomization, but this is limited in practice, leading to weak security.
We believe that randomization can still be an effective defense, but the degree of randomization needs to be significantly increased.
For this, we design {\em Fully Randomized Pointers} (\FRP{})---a method for randomizing as many pointer bits as possible without compromising binary compatibility.

\begin{example}[Fully Randomized Pointers]\label{ex:motive}
To illustrate the approach, consider the allocations ($p = \mathtt{malloc}(10)\texttt{;}~q = \mathtt{malloc}(10)$).
Possible values for $p$ and $q$ could be:

\vspace{-1em}
{
\setlength{\fboxsep}{0pt}
{\small
\setlength{\abovedisplayskip}{1em}
\setlength{\belowdisplayskip}{1em}
\begin{align}
p{=}\texttt{0x\colorbox{blue!20}{\strut 0000564745119010};} \tag{address} \label{eq:malloc} \\
q{=}\texttt{0x\colorbox{blue!20}{\strut 0000564745119020};} \tag{address} \\
p{=}\texttt{0x\colorbox{red!20}{\strut ac8415a209566010};} \tag{fully randomized} \label{eq:obfus} \\
q{=}\texttt{0x\colorbox{red!20}{\strut b5da178f9e40d020};} \tag{fully randomized}
\end{align}
}%
We assume an attacker who is aware of the value for $p$ but not for $q$,
e.g., with HeartBleed, the attacker knows $p{=}\texttt{buf}$, and must use $p$ to construct the address $q$ of the target object.
The (\ref{eq:malloc}) version is an example of an ordinary machine address returned by glibc \verb+malloc+.
Although the heap base address has been randomized using {\em Address Space Layout Randomization} (ASLR), the
relationship between the pointers is {\bf not} random and is predictable.
The attacker can easily infer the offset $k{=}q{-}p{=}16$ using basic knowledge of the internals of the version of \verb+malloc+ used by the program.

In contrast, almost all of the pointer bits in the (\ref{eq:obfus}) version have been randomized on a per-object basis, i.e., both $p$ and $q$ are essentially just random numbers.
Given the randomized value for $p$, what value for $k$ is needed to successfully overflow from $p$ into $q$?
Essentially, the attacker must solve $k{=}q{-}p$ where both $k$ and $q$ are {\em unknowns}, meaning that the difference $k$ is {\em undetermined}.
Given the high degree of randomization, brute forcing the specific $k$ is impractical, preventing the overflow.
}
$\qed$\end{example}

Although full pointer randomization is our goal, the question is ``how'' to do so.
The limitations of \ref{limit:x64} and \ref{limit:layout} (Sec.~\ref{sec:problem}) mean that a traditional object layout randomizer will not be sufficient.
We therefore propose a new design based on the simple premise that {\em pointers are not necessarily machine addresses}.
Rather, we take the view that a {\em pointer} \verb+p+ is an abstract entity that satisfies a set of basic properties:
\begin{enumerate}[leftmargin=*,noitemsep]
    \item \verb+p+ is associated with some underlying allocated object $O$;
    \item \verb+p+ identifies an {\em offset} in relation to the \emph{base} of $O$;
    \item the standard pointer operations (arithmetic, difference, dereference, etc.) can be applied to \verb+p+.
\end{enumerate}
A machine address is one possible pointer encoding, but other encodings are possible.
We hypothesize that most existing binary code is agnostic about the underlying pointer encoding; provided that these basic operations have the expected semantics.
Thus, we can replace the pointer encoding to introduce stronger memory safety, while retaining compatibility without recompilation.

Thus we design a {\em new} pointer encoding based on abstract $\langle\mathit{id},\mathit{offset}\rangle$ pairs, where $\mathit{id}$ is a unique identifier associated to a specific object $O$, and $\mathit{offset}$ is the explicit byte offset within $O$.
We instantiate the standard pointer operations (arithmetic, dereference, etc.), and show how this is sufficient for binary compatibility.
Importantly, the abstract encoding is:
\begin{enumerate}[leftmargin=*,noitemsep,label=(\alph*)]
\item \label{case:decouple} {\em decoupled} from the underlying object memory address; 
\item \label{case:obfus} both $\mathit{id}$ and $\mathit{offset}$ are {\em strongly} randomized---allowing for the relationship between pointers to be obfuscated and protecting against bypass attacks
\end{enumerate}
By ``decoupling'' the pointer encoding from the underlying memory address, the limitations of \ref{limit:x64} and \ref{limit:layout} are also no longer of concern---a key difference between our approach and existing heap randomizers (DieHard(er)~\cite{berge06diehard, gene10dieharder}).
Furthermore, even (partial) knowledge of object layout can no longer be exploited (\capLayout), and
the number of retries necessary for a brute-force style attack will be intractable.

The final challenge is performance.
In general, software implementations of memory safety tools have high overheads and are far too slow for widespread practical use.
Instead, we focus on a hardware implementation.
Since pointers are not represented as machine addresses, dynamic pointer decoding is necessary.
Furthermore, the mapping between pointers and addresses can change at any time as objects are \verb+malloc+'ed and \verb+free+'ed.

\begin{figure}[tb]
    \centering
    \includegraphics[scale=2]{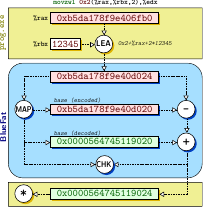}
    \caption{Example pointer decoding.}
    \label{fig:example}
\end{figure}

\subsection{Detailed Example}
We implement {\em Fully Randomized Pointers} (\FRP{}) by intercepting memory dereference operations and dynamically decoding pointer values.
A detailed example of on-the-fly pointer decoding is illustrated in Figure~\ref{fig:example}.
Here, we consider an example instruction with a (randomized) encoded pointer stored in \verb+%rax+.
The {\em Load Effective Address} (\textsf{LEA}) operation evaluates the memory operand \verb+0x2(%rax,%rbx,2)+, which corresponds to the pointer arithmetic
$(\texttt{0x2}+\texttt{\%rax}+2{\times}\texttt{\%rbx})$.
This operation is executed as normal under the standard \verb+x86_64+ semantics, but this time generates a derived encoded pointer rather than an address.
The dynamic pointer decoder is inserted between the (\textsf{LEA}) and dereference (\verb+*+) operations.
Here, the encoded access pointer is {\em mapped} (\textsf{MAP}) into an (encoded) base pointer ($q$ from Example~\ref{ex:motive}), and a (decoded) base memory address.
The difference (\verb+-+) between the (encoded) access pointer and base pointers is then added (\verb-+-) to the (decoded) base to yield the final (decoded) memory address.
The final address is dereferenced (\verb+*+) as per normal, before being discarded.
As such, only fully randomized (encoded) pointers are visible to the program, and these have no observable relation to the underlying memory address.

In addition to pointer decoding, our approach also checks for memory safety.
For {\em use-after-free} errors, our approach ensures that encoded pointers are very unlikely to be reused over the lifetime of the program.
This means that dangling pointer access will be detected by a missing entry in the (\textsf{MAP})---even when the underlying memory has been reallocated.
To check for {\em object-bounds} errors, (\textsf{MAP}) also stores object {\em metadata} (size+base address) which is checked before access (\textsf{CHK}):
\vspace{-1em}

\begin{align}
[\mathit{base}, \mathit{base}{+}\mathit{size})~\cap~[\mathit{lb},\mathit{ub})~=~[\mathit{lb},\mathit{ub})
    \tag{\textsf{CHK}}
\end{align}
\noindent
Here, $\mathit{base}$ is the object base address, $\mathit{size}$ is the object size, and $[\mathit{lb},\mathit{ub})$ are the bounds of the memory access.
Assuming that the object in Figure~\ref{fig:example} is of size 10 bytes (Example~\ref{ex:motive}), the resulting memory access is within bounds and thus is allowable.
Otherwise, if the address is out-of-bounds, our approach will detect this case and take protective action before the dereference operation (\verb+*+).

Finally, we must also consider the tricky case where the base address \verb+%rax+ is itself invalid.
For example, given the valid pointer \verb+p+ from Example~\ref{ex:motive}, an attacker could (in theory) pass the {\em invalid} (out-of-bounds) pointer $\texttt{p}'{=}\texttt{p}{+}k$ into \verb+%rax+.
This is possible since pointer arithmetic is never checked to avoid {\em false positives} at the binary level.
If the invalid $\texttt{p}'$ were to have the same value as a valid pointer $\texttt{q}$, then the invalid access will not be detected by either the (\textsf{MAP}) or (\textsf{CHK}) checks---effectively {\em bypassing} the defense.
Indeed, other memory safety tools, such as Valgrind~\cite{nethercote2007valgrind} and DrMemory~\cite{bruening2011drmem}, are vulnerable to precisely this kind of bypass attack. 
Essentially, the attacker needs to find the correct value for $k$, which is easy to do with default memory allocators that allocate objects within the same memory region.
In contrast, \FRP{} protects this case using a combination of pointer decoupling and randomization.
Namely, under our running example, the attacker must generate $\texttt{p}'$ from $\texttt{p}$ by finding the very specific value $k{=}672727314255736848$ out of ${\sim}2^{52}$ possibilities.
The attacker cannot exploit basic memory layout knowledge (\capLayout) since pointer values are decoupled.
Furthermore, the high degree of randomization means that the number of retries (\capRetry) necessary will easily exceed any practical attacker resource.

The problem of invalid pointers is also closely related to \emph{pointer provenance}~\cite{memarian2019ptr}, i.e., the relationship between pointer values and the underlying object that the object pointer references.
Two pointers may have the same value but can have different provenance (i.e., reference different objects), as shown with (invalid) $\texttt{p}'$ and valid $\texttt{q}$ in the example above.
For compatibility, tools based on memory poisoning (ASAN, Valgrind, etc.) do not track pointer provenance and are vulnerable to bypass attacks using invalid pointers.
Some memory error defenses, such as those based on {\em fat pointers}~\cite{jones1997bounds} or {\em capabilities} \cite{cheri}, explicitly track object metadata for provenance.
However, the explicit metadata tracking is generally not binary compatible and requires recompilation.
\FRP{} is effectively a form of {\em implicit} provenance tracking under a probabilistic guarantee, as different provenances correspond to very different (randomized) pointer values, and cannot easily be confused.
Unlike explicit tracking, \FRP{} does not require special handling of pointer arithmetic---thus preserving the {\em Application Binary Interface} (ABI) and achieving strong binary compatibility.
Since \FRP{} values are ordinary machine words, even advanced pointer idioms, such as XOR-linked lists, are fully supported.
Such idioms tend to break explicit metadata tracking.
Next, we detail our design:
pointer decoupling and randomization.

%%%%%%%%%%%%%%%%%%%%%%%%%%%%%%%%%%%%%%%%%%%%%%%%%%%%%%%%%
\section{Decoupled Pointers}\label{sec:basic}

We propose a {\em decoupled pointer} encoding that is (1) binary compatible, and (2) independent of the underlying memory layout, thereby defending against (\capLayout).
{\em Fully Randomized Pointers} (\FRP{}) will be built as an extension of decoupled pointers.

\newcommand{\reqAny}[0]{\textsf{Rq.*}}
\newcommand{\reqArith}[0]{\textsf{Rq.\textls[-10]{Arith}}\xspace}
\newcommand{\reqCast} [0]{\textsf{Rq.\textls[-10]{Word}}\xspace}
\newcommand{\reqDeref} [0]{\textsf{Rq.\textls[-10]{Deref}}\xspace}
\newcommand{\reqAlign} [0]{\textsf{Rq.\textls[-10]{Align}}\xspace}

\subsection{Pointer Requirements}
For compatibility, the pointer encoding must satisfy a set of basic requirements assumed by most binary code.
Here, each allocated object $O$ is uniquely identified by a \emph{base pointer} \verb+p+ returned by the underlying allocator.
The main pointer requirements (\reqAny) are:
\begin{enumerate}[leftmargin=2cm,noitemsep]
    \item[\hfill\textbf{\reqArith})] 
    $\texttt{p}_i{=}\texttt{p}{+}i$ points to the $i^{th}$ byte in $O$; and 
    $\texttt{p}_j{=}\texttt{p}_i{+}k$ (and $k{=}\texttt{p}_j{-}\texttt{p}_i$) for $k{=}(j{-}i)$.
    %\\%
    \item[\hfill\textbf{\reqCast})] $\texttt{p}_i$ is representable as a (64 bit) machine word.
    \item[\hfill\textbf{\reqDeref})] $\texttt{p}_i$ can be dereferenced to access $n$ consecutive $O$ bytes pointed to by $\texttt{p}_i..\texttt{p}_{i+n-1}$.
    \item[\hfill\textbf{\reqAlign})] $(\texttt{p}~\texttt{\&}~\texttt{0xFFF})$ must equal the alignment of $O$.
\end{enumerate}
A more detailed treatment of pointer requirements can be found in~\cite{memarian2019ptr}.
The (\reqAlign) requirement is often necessary for (binary) compatibility, where the binary determines object alignment by inspecting the pointer value.\footnote{
Most programs are compatible with 16-byte alignment ($\texttt{0xF}$).
Some programs as sensitive to page alignment ($\texttt{0xFFF}$), when interacting with system calls (e.g., \texttt{mprotect}) where page alignment is required.}
Many real-world programs also relax some of the requirements.
For example, many \verb+C+/\verb_C++_ programs (and most binaries) relax (\reqArith), allowing for arbitrary $k$ values outside the bounds of $O$.
In the case of \verb+C+/\verb_C++_, this is technically {\em undefined behavior} (\cite{iso2018cstd} \S 6.5.6 \P 8), but is nevertheless relied upon by many programs~\cite{duck2022redfat}.
Furthermore, many binaries relax (\reqDeref) for out-of-bounds \emph{reads}, provided that (1) the memory access does not cause a page fault (\verb+SIGSEGV+), and (2) any out-of-bounds data is discarded by subsequent execution.
For performance reasons, many \verb+glibc+ string handling functions (\verb+strlen+, \verb+strcpy+, etc.) use SIMD instructions that may read (then discard) out-of-bounds data.
We claim that any pointer encoding satisfying the relaxed (\reqAny) will be compatible with most existing binary code.

\subsection{Decoupled Pointer Design}

Our approach refactors pointers into $\langle\mathit{id},\mathit{offset}\rangle$ pairs:
\begin{itemize}[leftmargin=*,noitemsep]
    \item[-] $\mathit{id}$ is an {\em object identifier} that uniquely determines $O$; and 
    \item[-] $\mathit{offset}$ represents the byte offset from the base of $O$.
\end{itemize}
With this abstract encoding, a memory allocator returns a pointer of the form $\langle\mathit{id},\mathit{zero}\rangle$, where $\mathit{id}$ is a freshly minted (never-used-before) identifier, and $\mathit{zero}$ represents an offset of $0$ bytes.
Under the abstract encoding, the $\mathit{id}$ does not depend on the machine address---rather, objects are associated with unique identifiers that are never reused, even when the underlying memory is \verb+free+'ed and reallocated.
For this, identifiers can be assigned sequentially ($\#1$, $\#2$, $\#3$, $...$),
meaning that a dangling pointer can never be confused with a valid pointer to a reallocated object.
We also use the following definition of pointer arithmetic:
{
\setlength{\abovedisplayskip}{0.3em}
\setlength{\belowdisplayskip}{0.3em}
\begin{align}
    \langle\mathit{id},\mathit{offset}\rangle{+}k =
    \langle\mathit{id},\mathit{offset}{+}k\rangle
    \label{eq:arith}
    \tag{\textsc{Arith}}
\end{align}
}%
That is, pointer arithmetic only affects the $\mathit{offset}$,
meaning that an out-of-bounds abstract pointer can never be confused with a valid pointer to another object.
The object identifier ($\mathit{id}$) is idempotent with respect to reallocation and pointer arithmetic, resolving (in principle) all potential ambiguity between invalid and valid pointers.

\myparagraph{Flattening}
To satisfy (\reqCast), we {\em flatten} the encoding as:
\begin{Verbatim}
 struct { int64_t id:40; int64_t offset:24; };
\end{Verbatim}
Here $\mathit{offset}$ bitfield occupies the $m{=}\log_2(M)$ least significant bits (lsb), and the $\mathit{id}$ occupies the remaining $64{-}m$ most significant bits (msb), where $M$ is based on the default glibc \verb+malloc+ threshold\footnote{See the \texttt{mallopt} manpage.}     
$M{=}\texttt{MMAP\_THRESHOLD\_MAX}{=}16\mathit{MB}$.
Note that the $\mathit{offset}$ is stored within the lsb, meaning that integer encoding is consistent with (\ref{eq:arith}) for small values of $k$, satisfying (\reqArith).
Finally, to satisfy (\reqAlign), we use $\mathit{zero}{=}(\mathit{addr}~\texttt{\&}~\texttt{0xFFF})$, thereby
preserving the page offset (12 lsb) of the machine address ($\mathit{addr}$).

\myparagraph{Pointer Dereference}
The final requirement is pointer dereference (\reqDeref).
To address this, our approach is to intercept all memory accesses 
and to decode pointers ``on-the-fly'', as illustrated in Figure~\ref{fig:example}.
Essentially, the pointer dereference $(\texttt{*}\texttt{p})$ operation is replaced with $\texttt{*}\decodePtr(\texttt{p})$, where $\decodePtr$ maps encoded pointers to the corresponding machine address, defined as follows:
\begin{align*}
    \mathit{decode}(\texttt{p}) = \left\{ 
    \begin{array}{ll}
        \mathcal{M}[\mbox{\texttt{p}}.\mathit{id}].\mathit{base} + \mbox{\texttt{p}}.\mathit{offset} & \mathit{isEncoded(\mbox{\texttt{p}})} \\
        \mbox{\texttt{p}} & \mbox{otherwise} \\
    \end{array}
    \right.
\end{align*}
Here, $\isEncoded$ determines whether a pointer is encoded or not.
For the \verb+x86_64+, we can assume that a non-zero value for any of the 16 most significant bits will indicate an encoded pointer.
The mapping $\mapping$ is an {\em associative map} that maps identifiers of allocated objects to information about that object (also see \textsf{MAP} from Figure~\ref{fig:example}).
Here, $(\mapping[\mathit{id}].\mathit{base})$ is the {\em base machine address} of the object with $\mathit{id}$.
Initially, the map is empty ($\mapping{=}\emptyset$).
As execution proceeds, $\mapping$ is updated each time an object is (de)allocated, using wrappers over standard memory allocation functions. 

\subsection{Discussion}
Flattened decoupled pointers satisfy the relaxed requirements (\reqAny), and are a practical encoding that is compatible with existing binary code (without recompilation), assuming that pointers can be decoded on-the-fly.
Although the decoupled pointer encoding is compatible, it by itself only provides weak protection against bypass attacks.
Namely, (1) the $\mathit{offset}$ bitfield may {\em overflow} into the $\mathit{id}$, meaning that $\mathtt{q}{=}\mathtt{p}{+}k$ still has a solution, and (2) the $\mathit{id}$ bitfield is allocated in sequence, meaning that it may have a predicable value.
We shall address security using full randomization.

\section{Fully Randomized Pointers}\label{sec:obfus}

We use the decoupled pointer encoding from Section~\ref{sec:basic} as the basis for {\em Fully Randomized Pointers} (\FRP{}).
For binary compatibility, all pointers must be represented as 64 bit integers (\reqCast), meaning that the 
object identifier bitfield ($\mathit{id}$) is vulnerable to modification, and this is difficult to prevent directly.
Even if all pointer arithmetic were to be fully instrumented (e.g., to detect bitfield overflows), this would break compatibility with binary code that creates deliberate out-of-bounds pointers (relaxed \reqArith).
Instead, we use an indirect mitigation in the form of full randomization.

\subsection{Decoupled Pointer Randomization}

The idea is to obfuscate the relationship between pointers to different objects, meaning that the necessary value for $k$ to construct an out-of-bounds pointer (\capInvalid) becomes undetermined.
Similar protection is afforded to dangling pointers, thereby hardening against bypass attacks in general.
Unlike existing randomization-based defenses, such as DieHard(er)~\cite{berge06diehard, gene10dieharder}, our baseline pointer encoding is decoupled from the underlying memory address, meaning that we randomize pointer bits without affecting the layout of objects within memory.
Decoupled pointer randomization is relatively straightforward: the \emph{identifier} ($\mathit{id}$) and \emph{offset} bitfields are randomized independently, as discussed below.

\myparagraph{Identifier randomization}
Identifiers can be generated at random with a collision-check:
\begin{align*}
    \assignId() = \left\{\begin{array}{ll}
        \mathit{id} & \mathit{id} \leftarrow \mathit{getId}(), \mathit{id} \not\in \mathrm{dom}~\mapping \\
        \assignId() & \text{otherwise}
        \end{array}\right.
\end{align*}
Here, $\mathit{getId}()$ is a {\em Cryptographically Secure Pseudo Random Number Generator} (CSPRNG) source.
Since different objects will be assigned very different (random) ids, there is no predictable relationship between their respective values, defending against (\capInvalid).

\myparagraph{Offset randomization}
It is also possible to randomize the $\mathit{offset}$.
This exploits the observation that the $\mathit{zero}$ can be any arbitrary value subject to the following constraints:
\begin{enumerate}[leftmargin=*,noitemsep]
    \item \label{case:size} 
    $(\mathit{zero} + \mathit{size})$ cannot overflow the offset bitfield, and
    \item \label{case:lsb}
          $(\mathit{zero}\texttt{ \& 0xFFF})$ must match the decoded machine base pointer to satisfy (\reqAlign).
\end{enumerate}

\myparagraph{Additional optional randomization}
It is also possible to (optionally) randomize the page offset and the default 16 byte malloc alignment, allowing for an additional 12 bits of randomization.
However, this will be incompatible for some binaries, so it is disabled by default.

\myparagraph{Sensitive data}
An attacker may also attempt to bypass the defense by reading sensitive data, such as the mapping $\mapping$, CSPRNG-pa\-ram\-et\-ers, or the heap itself (via a machine address).
For our analysis, we assume that sensitive data is stored in a dedicated memory region that is not accessible by the program, i.e., inaccessible or not addressable.
For a software implementation, this can be achieved with additional instrumented checks.
For hardware, this can be achieved by storing sensitive information in an inaccessible memory region.
Note that the heap itself is considered sensitive data.
This means that heap objects are only accessible via {\FRP{}}s and not regular addresses.

\subsection{Security Analysis}
We model two single-word objects pointed to by encoded pointers \verb+p+, \verb+q+.
We assume that the attacker knows the value of \verb+p+, but not the value of the target pointer \verb+q+, which can take any value in the \FRP{} space.
The attacker can construct (\capInvalid) and dereference (\capDeref) invalid pointers.
Furthermore, the attacker is able to retry failed attacks (\capRetry).
The attacker is adversarial and can attempt any attack allowable under the threat model.

We can quantify the probability of isolated attack attempts succeeding as one in the following value:
\begin{align}
(2^{|\mathit{id}|}-1-2^{|\mathit{id}|-16}) \times 2^{|\mathit{offset}|} \times 2^{|\mathit{page}|} \times 2^{|\mathit{align}|} \tag{\textsc{Security}} \label{eq:security}
\end{align}
Here, $|\mathit{id}|$ is the number of randomized bits in the $\mathit{id}$, $|\mathit{offset}|$ is the number randomized bits in the $\mathit{offset}$, $|\mathit{page}|$ is the number of randomized bits in the page offset (optional), and $|\mathit{align}|$ is the remaining number randomized bits that is otherwise reserved for \verb+malloc+ alignment (optional).
Here (\ref{eq:security}) accounts for $2^{|\mathit{id}|}$ potential randomized identifiers, less one already consumed by \verb+p+, less ($2^{|\mathit{id}|-16}$) identifiers with zero 16 msb (reserved for non-encoded addresses).
For our baseline, we assume $|\mathit{id}|{=}40$, $|\mathit{offset}|{=}12$, and $|\mathit{page}|{=}|\mathit{align}|{=}0$ which evaluates to \textbf{52 bits}---significantly stronger than the current state-of-the-art~\cite{lemay2021c3, saileshwar2022heapcheck, li2022pacmem}.
This analysis conservatively assumes the attacker is aware of the memory layout (\capLayout) and can therefore guess the lower 12 bits of the target \verb+q+.
By relaxing these assumptions, we have that $|\mathit{page}|{=}8$ and $|\mathit{align}|{=}4$, meaning that the effective entropy increases to \textbf{60 bits} and \textbf{64 bits} respectively.
We now consider some possible attack scenarios:

\myparagraph{Network server reset}
The attacker attempts to dereference an invalid pointer, hoping to hit the target \texttt{q}.
For an incorrect guess, the program terminates, but the attacker can retry under (\capRetry).
Under the baseline entropy of $52$ bits, the attacker will be required to make an average of $2^{52-1}$ guesses before a successful attack.
Assuming the attacker targets a network server that resets after a crash, the network bandwidth for the TCP handshakes ($3{\times}40$ bytes) alone would exceed \textbf{240 petabytes}.
As such, this is not a practical attack.

\myparagraph{Speculative execution}
Here, the attacker dereferences invalid pointers \emph{speculatively}, and uses side channels (e.g., timing) to determine if the guess was valid or not.
The advantage of speculation is that it does not terminate the program if the attempt fails.
However, this also requires a more complicated setup, and may require sampling of noisy side channels.
For our analysis, 
we use the PACMan attack~\cite{ravi2022pacman}, a practical speculative execution attack against ARM {\em Pointer Authentication Codes} (PAC).
We use the estimate of 2.69 milliseconds per attack attempt (\cite{ravi2022pacman}~\S~8.2), which can exhaust all 16 bit PAC values in 2.94 minutes, and all 24 bit values in 12.54 hours.
At the same rate, exhausting all 52 bit \FRP{} values would take \textbf{384154 years}.
Again, this is not a practical attack.

\myparagraph{Heap spray}
The attacker may induce the program to create multiple $n$ targets (heap spray), rather than assuming a single target ($n{=}1$).
This is another form of (\capRetry), where $n$ is the number of attack attempts.
The disadvantage of this approach is that each allocated target consumes memory, which imposes a practical bound on the size of $n$.
Furthermore, memory size is orders of magnitude smaller than the size of the \FRP{} space, meaning that heap sprays have a negligible impact on security.
For example, suppose the attacker allocated $2^{32}$ targets, which by itself would consume considerable memory.
However, the effective security of \FRP{} would be essentially unchanged: $2^{52}{-}2^{32}{\approx}2^{52}$.
The attacker may also attempt to allocate objects to reduce the size of the (unallocated) \FRP{} space, making future allocations more predictable.
However, this runs into the same problem: the \FRP{} space is so large that it will not be meaningfully affected by such attacks, using the same reasoning as above.

\section{Evaluation}\label{sec:eval}
We have implemented a software implementation (\bluefat) of {\em Fully Randomized Pointers} (\FRP{}) on top of the Pin {\em Dynamic Binary Instrumentation} (DBI) framework~\cite{luk05pin} and a hardware simulation (\greenfat) on top of gem5~\cite{gem5simulator}.
The software implementation primarily aims to comprehensively test the security and compatibility of \FRP{} on real-world \verb+x86_64+ binaries without recompilation.
The hardware simulation aims to evaluate the performance potential of \FRP{} under the assumption of a native implementation in hardware.

In both cases, the \FRP{} wraps (using \verb+LD_PRELOAD+) the standard \verb+glibc+ memory allocation functions with full pointer randomization, and intercepts all memory dereferences to dynamically decode pointers into the corresponding memory address.
The implementation also checks for memory errors and will abort the program on use-after-free or out-of-bounds write.
For reads, the implementation will zero any out-of-bound byte (similar to {\em failure-oblivious computing}~\cite{rinard2004fail}), retaining compatibility with the relaxed (\reqDeref) while also preventing exploitation in the form of information disclosure.
In terms of look-and-feel, the software implementation (\bluefat) is similar to other DBI-based memory error tools, such as Valgrind~\cite{nethercote2007valgrind} and DrMemory~\cite{bruening2011drmem}.
However, unlike these tools, the \FRP{} methodology is designed to harden against bypass attacks (see comparison in Table~\ref{tab:sanitizers}).

All experiments are run on an Intel Xeon Gold
6242R CPU (clocked at 3.10GHz) with 376GB of RAM.
Each benchmark binary is obtained by compiling using \verb+gcc+/\verb_g++_/\verb+gfortran+ compiler version 9.4.0 under (\verb+-O2+).
The system runs on Ubuntu 20.04 (LTS).
We evaluate \bluefat for security and compatibility, and \greenfat for performance.

\subsection{Security}

\FRP{} is primarily optimized for security and resistance to bypass attacks.
We evaluate \bluefat against prominent memory error defenses from Table~\ref{tab:sanitizers}.
For benchmarks, we use (in increasing order of difficulty):
\begin{enumerate}[leftmargin=*,noitemsep]
    \item Recent {\em Common Vulnerabilities and Exposures} (CVEs); 
    \item \textsc{RedFat} benchmarks (from~\cite{duck2022redfat}) including {\em non-inc\-re\-men\-tal}~\cite{duck2022redfat} CVEs and 480 Juliet (Jul.)~\cite{juliet} test cases;
    \item A {\em strong} adversarial attacker model capable of brute-force bypass attacks (see Sec.~\ref{sec:attacker}).
    This includes ported \textsc{RedFat} CVEs and the Table~\ref{tab:sanitizers} micro-benchmarks.
\end{enumerate}
We caution against drawing conclusions from the {\em quantity} of passing/failing tests.
For example, the \textsc{RedFat} benchmarks from~\cite{duck2022redfat} (which are used ``as-is'') are
evaluated w.r.t. Valgrind.
Rather, {\em any} failing test means that the tool is vulnerable to bypass attacks (see the {\em Bypass safe?} column).
Unlike the other tools, DieHard(er) implements error \emph{mitigation} rather than detection.
This is represented by a dash (-) and should not be interpreted as a negative outcome.
We evaluate all tools from Table~\ref{tab:sanitizers} for which implementations are publicly available and can run under our test setup.

\begin{table}[bt]
\setlength{\tabcolsep}{2.3pt}
\footnotesize
\centering
\caption{Security evaluation of defense techniques. 
\label{tab:security_evaluation}}
\begin{tabular}{|l|cccc|cccc|r|cccc|cc|c|}
\cline{2-16}
\multicolumn{1}{c|}{} & \multicolumn{4}{c|}{\em Recent} &
                   \multicolumn{5}{c|}{\textsc{RedFat} {\em Bench.}} &
                   \multicolumn{6}{c|}{\em Strong} &
                   \multicolumn{1}{c}{} \\
\cline{2-16}
\multicolumn{1}{c|}{} & \multicolumn{4}{c|}{CVEs} &
                   \multicolumn{4}{c|}{CVEs} & \multicolumn{1}{c|}{Jul.} &
                   \multicolumn{4}{c|}{CVEs} &
                   \multicolumn{2}{c|}{\textls[-30]{Micro}} &
                   \multicolumn{1}{c}{}
                   \\
\cline{2-17}
\multicolumn{1}{c|}{\emph{Defense}}
                 & \rot{\em cve-2023-29584~~}
                 & \rot{\em cve-2023-25221~~}
                 & \rot{\em cve-2023-27249~~}
                 & \rot{\em cve-2023-25222~~}
                 & \rot{\em cve-2007-3476~~}
                 & \rot{\em cve-2012-4295}
                 & \rot{\em cve-2016-1903}
                 & \rot{\em cve-2016-2335}
                 & \multicolumn{1}{c|}{\rot{\em cwe-122-heap~}}
                 & \rot{\em cve-2007-3476*}
                 & \rot{\em cve-2012-4295*}
                 & \rot{\em cve-2016-1903*}
                 & \rot{\em cve-2016-2335*}
                 & \rot{$\mathit{OF}{+}\mathit{UF}$}
                 & \rot{$\mathit{UAF}$}
                 & \rot{\em Bypass safe?}
                 \\
\hline
\emph{ASLR}+\texttt{glibc} &
    \xmark & \xmark & \xmark & \xmark &\xmark & \xmark & \xmark & \xmark & {\color{red!50!black}0} & \xmark & \xmark & \xmark & \xmark & \xmark & \xmark & \xmark \\
\hline
ASAN~\cite{asan} &
    \cmark & \cmark & \cmark & \cmark & \cmark & \cmark & \cmark & \cmark & {\color{red!50!black}210} & \xmark & \xmark & \xmark & \xmark & \xmark & \xmark & \xmark \\
LowFat~\cite{duck16heap, duck17stack} &
    \xmark & \xmark & \xmark   & \cmark   & \cmark &  \xmark  & \xmark & \xmark  &   {\color{red!50!black}428} & \xmark & \xmark & \xmark & \xmark & \xmark & \xmark & \xmark \\
EffectiveSan~\cite{duck18effective} &
    \cmark & \cmark & \xmark   & \cmark   & \cmark & \cmark & \cmark & \cmark &   {\color{green!50!black}480} & \xmark & \xmark & \xmark & \xmark & \xmark & \xmark & \xmark \\
\hline
\emph{efence}~\cite{bruce1993electri} &
    \xmark & \cmark & \xmark & \cmark & \cmark   & \cmark   &  \cmark  &  \cmark  &  {\color{green!50!black}480}   &  \xmark & \xmark & \xmark & \xmark & \xmark    &  \xmark & \xmark \\
DieHard~\cite{berge06diehard} &
     - & - & - & - & - & - & - & - & - & \xmark & \xmark & \xmark & \xmark &  \xmark & \xmark & \xmark \\
DieHarder~\cite{gene10dieharder} &
     - & - & - & - & - & - & - & - & - & \xmark & \xmark & \xmark & \xmark &  \xmark & \xmark & \xmark \\
Valgrind~\cite{berge06diehard} &
     \xmark & \cmark & \cmark & \cmark & \xmark & \xmark & \xmark & \xmark & {\color{red!50!black}0} & \xmark & \xmark & \xmark & \xmark &  \xmark & \xmark & \xmark \\
DrMemory~\cite{bruening2011drmem} &
     \xmark & \cmark & \cmark & \cmark & \xmark & \cmark & \cmark & \xmark & {\color{red!50!black}476} & \xmark & \xmark & \xmark & \xmark &  \xmark & \xmark & \xmark \\
\textsc{RedFat}~\cite{duck2022redfat} &
     \cmark & \xmark & \xmark & \cmark & \cmark & \cmark & \cmark & \cmark & {\color{green!50!black}480} & \xmark & \xmark & \xmark & \xmark &  \xmark & \xmark & \xmark \\
\hline
\hline
\FRP{} (this work)&
     \cmark & \cmark & \cmark & \cmark & \cmark & \cmark & \cmark & \cmark & {\color{green!50!black}480} & \cmark & \cmark & \cmark & \cmark & \cmark & \cmark & \cmark \\
\hline
\multicolumn{1}{l}{} \\
\end{tabular} \\
\vspace{-1em}
\begin{tabular}{c}
({\cmark}) = error detected, ({\xmark}) = error not detected \\
(-) = {\em indeterminate} since the benchmark does not model a target object. \\
(Jul.) = Juliet benchmarks (480 tests); (*) = ported to strong attacker
\end{tabular}
\end{table}

The results are shown in Table~\ref{tab:security_evaluation}.
Here, we see that most existing defenses can detect ``ordinary'' CVEs (\emph{Recent}) during normal execution---i.e., 
without an attacker that is actively attempting to bypass the defense.
Such tools are useful for debugging or at least some modest attack surface reduction.
The \textsc{RedFat} benchmarks use {\em non-incremental} overflows which allow for the construction of pointers with attacker-controlled offsets ($p{+}k$).
This leads to mixed results on some defenses---the value chosen for $k$ may be sufficient to bypass common detection methods, e.g., memory poisoning.
Finally, we consider our ``strong'' threat model (\emph{Strong}).
For this, we port the \textsc{RedFat} CVEs to also include a proactive attacker able to do (\capAny), including being able to retry (brute force) failed attacks.
We also test the micro benchmarks from Sec.~\ref{sec:problem}, and an additional \emph{use-after-free} ($\mathit{UAF}$) benchmark:

{\small
\begin{Verbatim}[commandchars=\\\{\},codes={\catcode`$=3\catcode`^=7}]
  free(p); 
  \textbf{for} (i = 0; i < N; i++) \{malloc($n$); \textbf{attack}(p);\}
\end{Verbatim}
}

\noindent
Our tests use a maximum of 10,000 attempts.
All defenses fail except \tool, showing these defenses do not protect against strong attackers.
As pointers have been randomized, the number of attempts necessary for a successful attack is ${\sim}2^{51}$ 
providing strong security even against a capable attacker.
Table~\ref{tab:security_evaluation} also highlights the problem of incomplete error coverage due to uninstrumented code (e.g., shared libraries), as illustrated by CVE-2023-27249, which is missed by many tools.
Such problems are avoided by transparently checking {\em all} memory access, which is possible with a hardware implementation (\greenfat).

The Table~\ref{tab:security_evaluation} 
reinforces the results from Table~\ref{tab:sanitizers} (see Sec.~\ref{sec:problem}).
Namely, that existing defenses are effective against weak or non-existent attacker models (\emph{Recent}), have mixed results against less-weak attack models, and have essentially no resistance against strong attacker models (\emph{Strong}).
Given that the strong attacker model is not unrealistic,  this motivates a stronger memory error defense such as \FRP{}.

\begin{figure*}
\centering
\pgfplotsset{
    axisA/.style={
        ybar=0pt,
        ymin=-0.1,
        ymax=200,
        scaled ticks=false,
        xticklabels={
\texttt{\scriptsize perlbench},
\texttt{\scriptsize gcc},
\texttt{\scriptsize bwaves},
\texttt{\scriptsize mcf},
\texttt{\scriptsize cactuBSSN},
\texttt{\scriptsize lbm},
\texttt{\scriptsize omnetpp},
\texttt{\scriptsize wrf},
\texttt{\scriptsize xalancbmk},
\texttt{\scriptsize x264},
% \texttt{\scriptsize cam4},
\texttt{\scriptsize pop2},
\texttt{\scriptsize deepsjeng},
\texttt{\scriptsize imagick},
\texttt{\scriptsize leela},
\texttt{\scriptsize nab},
\texttt{\scriptsize exchange2},
\texttt{\scriptsize fotonik3d},
\texttt{\scriptsize roms},
\texttt{\scriptsize xz},
{\scriptsize Geo. Mean $*$},
        },
        ytick={0, 50, 100, 150, 200},
        yticklabels={
            {\scriptsize 0},
            {\scriptsize 50\%},
            {\scriptsize 100\%},
            {\scriptsize 150\%},
            {\scriptsize 200\%},
        },
        bar width=4.5pt,
        x tick label style={rotate=45,anchor=east,yshift=-2.5pt,xshift=3pt},
        width=\textwidth,
        height=4.25cm,
        xtick=data,
        xtick pos=left,
        ytick pos=left,
        major tick length=0.08cm,
        enlarge x limits={true, abs value=0.8},
        grid style={gray!20},
        grid=both,
        title style={yshift=-6pt},
        nodes near coords,
        every node near coord/.append style={
            xshift=-0.03cm,
            yshift=-0.08cm,
        },
        point meta=explicit symbolic,
        legend cell align={left},
        legend pos=north east,
        legend image code/.code={% 
            \draw[#1] (0cm,-0.1cm) rectangle (3pt,3pt);
        },
        legend columns=1,
        }
}

\pgfplotstableread[col sep=comma, header=false]{
0,	1.39,	15.16,	55.81,	100.00,
1,	3.46,	29.58,	180.69,	100.00,
2,	0.93,	10.38,	91.62,	100.00,
3,	1.50,	5.45,	99.77,	100.00,
4,	0.33,	5.52,	30.65,	100.00,
5,	0.95,	3.98,	87.00,	100.00,
6,	1.10,	8.43,	56.18,	100.00,
7,	3.26,	22.58,	118.19,	100.00,
8,	2.64,	25.37,	113.26,	100.00,
9,	2.96,	19.3,	141.16,	100.00,
% 10,	   0,	16.7,	  0,	      0.00, $\dagger $
10,	1.76,	20.55,	147.33,	100.00,
11,	0.38,	9.13,	39.26,	100.00,
12,	1.67,	17.4,	84.33,	100.00,
13,	0.73,	8.46,	54.46,	100.00,
14,	1.13,	7.79,	44.95,	100.00,
15,	0.56,	9.74,	52.35,	100.00,
16,	2.95,	15.81,	163.41,	100.00,
17,	1.56,	14.28,	103.79,	100.00,
18,	2.17,	5.14,	111.24,	100.00,
19,	1.43,	11.65,	87.70,	100.00,
}\dataset
\begin{tikzpicture}
    \begin{axis}[
        axisA
      ]
      \addplot[pattern color=red, pattern=north east lines] table[x index=0,y index=4, meta=5] \dataset;
      \addplot[fill=blue] table[x index=0,y index=3] \dataset;
      \legend{{\footnotesize Valgrind},
              {\footnotesize \tool}};
      \end{axis}
    after end axis/.append code={
  \draw[thin] (current axis.left of origin) -- (current axis.right of origin);
}
\end{tikzpicture}
\vspace{-3mm}
\caption{Normalized SPEC timings for Valgrind and \tool.
For (*) we exclude \texttt{cam4}, which runs under \tool but not Valgrind.
The results demonstrate perfect compatibility for \tool and \FRP{}--despite pointers being radically different.
The absolute slowdown for \tool is $11.19{\times}$, mainly due to DBI costs.
\label{fig:performance}}
\vspace{-6pt}
\end{figure*}

\subsection{Compatibility}
To test compatibility, we evaluate \bluefat on the complete SPEC2017 benchmark suite~\cite{bucek2018spec} under the \verb+ref+ workload.
There are 19 benchmarks: 12 \texttt{C} benchmarks, 5  \texttt{C++}, and 8 \texttt{Fortran} (non-exclusively).
The results are shown in Figure~\ref{fig:performance}, and represent the relative slowdown using Valgrind's Memcheck as a baseline.
To make the comparison fairer, we disable leak checking (\verb+–leak-check=no+) and undefined value errors (\verb+–undef-value-errors=no+), which speeds up Valgrind somewhat.
We compare against \tool that uses a software cache 
 size of 64KB (4096 entries, 16 bytes per entry) to reduce the number of times the associative map ($\mapping$) needs to be consulted.

We compare against Valgrind as a \emph{Dynamic Binary Instrumentation} (DBI)-based memory error detection tool that is binary compatible without recompilation.
We caution that compatibility (and not performance) is the primary goal of these experiments---i.e., can \FRP{} (as implemented by \bluefat) run the full SPEC2017 benchmark suite without error?
Our main performance result is based on a hardware simulation (\greenfat
see Section~\ref{sec:hardware}).

Binary compatibility is essential for a usable tool.
Despite using a radically different pointer encoding (i.e., \FRP{} replacing addresses with random numbers), \tool can successfully run the entire SPEC benchmark suite (Figure~\ref{fig:performance}) without any special (re)compilation or instrumentation, i.e., the binary itself is unaware of the hardening solution.
This validates the key premise of our design, namely: alternative pointer encodings are indeed practical, provided that a minimal set of pointer requirements (\reqAny) have been satisfied.
The results demonstrate that \FRP{} can be applied to any binary, including legacy binaries and libraries, without rebuilding the entire software tool chain.
We also tested \tool on a wide plethora of off-the-shelf binaries, including~(\cite{dinesh2020retro} Table~II {\em Real-World}), (\cite{duck20binary} Table 1), and many other programs/libraries.
We find that \tool is highly compatible with existing binaries---further validating our approach.

Other memory safety tools have mixed results, despite using the default pointer encoding.
Valgrind can run most benchmarks except \texttt{cam4} (which failed due to a segmentation fault under our tests).
We also tested DrMemory~\cite{bruening2011drmem}, with \texttt{deepsjeng} reporting a heap allocation failure, but with the other benchmarks running, albeit much slower than Valgrind.
While other DBI-based memory error detection tools can also achieve reasonable compatibility, only \FRP{} achieves the strong bypass resistance and security guarantees necessary for hardening applications.

\subsection{Performance}\label{sec:hardware}

In this section, we demonstrate a proof-of-concept hardware-accelerated version of \FRP{} (\greenfat{}). 
Our goal with \greenfat{} is to propose a hardware-software co-designed solution that significantly reduces the overheads seen when using \FRP{}. It does this by storing a mapping from the \FRP{} value to the virtual address and bounds metadata in a small, fast-to-access cache. This accelerates the translation of an \FRP{} and allows for the hardware to quickly check recently used \FRP{} values for the translations in this cache. In the relatively rare case of a cache miss, we access the translation data from DRAM and cache it for future accesses.

In the rest of this section, we describe the methodology used for application representative generation, detailed simulator setup, and provide an analysis.

\subsubsection{Methodology}

We model our \FRP{} implementation, \greenfat{}, using the gem5~\cite{gem5simulator,lowepower2020gem5} simulator. The gem5 simulator is a popular cycle-level processor simulation methodology.
We model a modern processor system with a full cache hierarchy, and a detailed out-of-order CPU execution model (See Table~\ref{tab:hardware_configuration} for details) in Syscall Emulation mode. We then augment this system with the \greenfat{} cache to build a fast-to-access memory that tracks recent object-to-bounds translations, minimizing overheads and allowing for fast lookups of recently-used \FRP{} mappings.

The \greenfat{} cache is an 8-way associative structure modeled with $\{128, 512, 1024, 4096\}$ entries, and each entry stores the virtual address bounds information for each object (assumes 16 bytes of space per entry, same as \bluefat). These caches have a total capacity of \{2\,KiB, 8\,KiB, 16\,KiB, 64\,KiB\} respectively. Entries are added to the \FRP{} mappings through our memory allocation wrappers (\verb+malloc()+, \verb+free()+, etc.) that do not require recompilation of the binaries under test.
The \greenfat{} cache is also connected to the Last-Level Cache (LLC) to minimize miss latency.

The evaluation is done using the working subset of C/C++ applications\footnote{For the setup used to build representatives, the RISC-V compiler used (LLVM 10.0) does not support Fortran applications, and the gem5 version used has many unimplemented syscalls, limiting application support~\cite{hajiabadi2021noreba}.}
from the SPEC CPU2006 benchmark suite\footnote{We switch to SPEC CPU2006 for compatibility with our gem5 setup.} 
built using LLVM 10.0 compiler with a RISC-V backend.
As running full applications in detail %ed timing mode 
can take months to simulate in gem5~\cite{sabu2022looppoint}, 
we use application representatives that accurately represent the original workload behavior, a standard methodology for performance benchmarking~\cite{hamerly2005simpoint}.
To generate these representatives, we use functional simulation of gem5 (AtomicCPU mode) to collect the Basic Block Vectors (BBVs) required to cluster the regions to generate the one billion (1\,B) instruction SimPoint~\cite{hamerly2005simpoint} representatives. 
The full reference input sets were used (largest input sets), running both the (i) baseline gem5, and the (ii) \greenfat configuration. We collect performance and cache statistics, and detail those results later in this section. 
All object statistics were generated using the 1\,B representative region execution.

\subsubsection{\greenfat{} Implementation Details.}

In this section, we will present an overview of the steps taken by the hardware-software co-designed system to implement \FRP{}s.
First, (1) upon heap memory allocation, we allocate a mapping from a newly minted \FRP{} to a virtual address region. This is done at the time of the \verb+malloc()+ function call. Next, (2) on access by a standard load or store instruction, the hardware checks for the presence of an \FRP{}, and, when identified, (2a) looks up the \FRP{} in the \greenfat{} cache to determine the object base and bounds. On a hit, the access continues as normal, but on a miss (2b), the \greenfat{} cache reads the object map ($\mapping$) stored in DRAM (and cached using the LLC), and copies the corresponding entry into the \greenfat{} cache for later use.
The hardware next (3) checks for validity of the address, and if (3a) valid, proceeds to access the L1-D cache using the virtual address translated from the \FRP{}.
Note that metadata from the \greenfat{} cache needs to be checked 
before initiating an L1-D cache lookup via the translated virtual address. If invalid (3b), the hardware will raise a memory access exception (similar to an invalid memory access), and the OS will raise an exception.

\begin{table}[t]
\caption{Configuration for gem5 simulation. Acronyms are: Out-of-Order CPU (OoO), Load Queue (LQ), Store Queue (SQ), Reorder Buffer (ROB), Least-Recently Used (LRU).
}
\label{tab:hardware_configuration}
\begin{tabular}{@{} ll @{}}
\toprule
{Component} & {Description} \\ 
\midrule
{CPU} & {OoO @ 1\,GHz, 32 LQ, 32 SQ} \\
{} & {192 ROB entries} \\
\greenfat{} Cache & 8-way LRU, 16\,B line size \\
{} & 2 cyc lat  \\
L1I/D Cache & 32\,KiB, 2-way LRU \\
{} & 2 cyc lat, 64\,B line size \\
LLC Cache & 256\,KiB, 8-way LRU \\
{} & 20 cyc latency, 64\,B line size \\
DRAM & tCL 13.75\,ns, tRCD 13.75\,ns \\
{} & tRP 13.75\,ns (approximately 80 cycles) \\
\bottomrule
\end{tabular}
\end{table}

\begin{table*}[t]
\centering
\small
\caption{Benchmark slowdown (\%) and cache miss rates (\%) for gem5 simulations of our \greenfat{} implementation across different cache sizes (4096, 1024, 512, and 128 entries of an 8-way associative cache). Slowdown is calculated by comparing the benchmark simulated runtime under \greenfat{} with baseline gem5. Object data for the representative regions is also listed. Entries with a hyphen (-) indicate a cache miss rate of less than 0.01\%. Count is the number of unique objects accessed in the representative region. The $\mu$ symbol represents the mean object size (in KiB), and $|A|$ is the number of accesses, in millions.}
\label{table:benchmarkgem5data}
\setlength{\tabcolsep}{3.5pt}
\begin{tabular}{@{} l rrrrrrrrrrr @{}} 
\toprule
\multirow{2}[1]{*}{Benchmark} & \multicolumn{4}{c}{Slowdown (\%)} & \multicolumn{4}{c}{Cache Miss Rate (\%)} & \multicolumn{3}{c @{}}{Object Data} \\
\cmidrule(lr){2-5} \cmidrule(lr){6-9} \cmidrule(l){10-12}
& {4096} & {1024} & {512} & {128} & {4096} & {1024} & {512} & {128} & Count &  $\mu$ (KiB) & $|A|$ (M) \\
\midrule

{401.bzip2}      & {3.35}  & {3.35}  & {3.35}  & {3.35}  & {-}     & {-}     & {-}     & {-}     & {7}       & {7,329.8}   & {171.4}  \\ 
{403.gcc}        & {2.01}  & {2.05}  & {2.24}  & {3.90}  & {0.03}  & {0.04}  & {0.06}  & {0.48}  & {8,901}   & {2.5}       & {230.9}  \\ 
{429.mcf}        & {4.79}  & {4.79}  & {4.79}  & {4.79}  & {-}     & {-}     & {-}     & {-}     & {3}       & {190,698.8} & {1043.7} \\ 
{433.milc}       & {1.94}  & {1.94}  & {1.94}  & {1.94}  & {-}     & {-}     & {-}     & {-}     & {40}      & {11,798.5}  & {357.1}  \\ 
{445.gobmk}      & {0.22}  & {0.22}  & {0.22}  & {0.22}  & {-}     & {-}     & {-}     & {-}     & {8}       & {1,487.6}   & {18.5}   \\ 
{462.libquantum} & {0.77}  & {0.77}  & {0.77}  & {0.77}  & {-}     & {-}     & {-}     & {-}     & {1}       & {8,193.0}   & {253.0}  \\ 
{464.h264ref}    & {0.38}  & {0.41}  & {0.62}  & {1.28}  & {0.00}  & {0.01}  & {0.05}  & {0.13}  & {1,164}   & {1.8}       & {292.7}  \\ 
{470.lbm}        & {3.98}  & {3.98}  & {3.98}  & {3.98}  & {-}     & {-}     & {-}     & {-}     & {2}       & {52,344.8}  & {311.2}  \\ 
{473.astar}      & {15.08} & {16.04} & {16.43} & {17.76} & {19.19} & {20.68} & {22.45} & {24.33} & {192,298} & {0.5}       & {63.0}   \\ 
{483.xalancbmk}  & {7.50}  & {15.26} & {34.39} & {67.33} & {0.60}  & {1.44}  & {2.97}  & {5.18}  & {66,444}  & {0.6}       & {231.0}  \\
\midrule
GEOMEAN        & {\textbf{3.92}}  & {4.74} & {6.44}  & {9.17}  & {-}     & {-}     & {-}     & {-}     & {-}       & {-}         & {-} \\

\bottomrule
\end{tabular}
\end{table*}

\subsubsection{Analysis}

We assess \greenfat{} cache statistics and the impact on performance over four different cache-sizes — 2\,KiB, 8\,KiB, 16\,KiB, and 64\,KiB corresponding to 128 entries, 512 entries, 1024, and 4096 entries, respectively. Table~\ref{table:benchmarkgem5data}
displays the \greenfat{} cache miss rates and performance slowdown of \greenfat{} compared to the baseline gem5 implementation.
As \greenfat{} stores mappings in the \greenfat{} cache on a per-object basis,
the miss rate remains low (<0.01\%) for six of the benchmarks (\verb+401.bzip2+, \verb+433.milc+, \verb+429.mcf+, \verb+445.gobmk+, \verb+462.libquantum+, and \verb+470.lbm+).
Object counts range from just one for \verb+462.libquantum+, to 40 objects in \verb+433.milc+.
Thus, workloads with a lower number of objects have low cache-sensitivity.

Contrary to that, benchmarks such as \verb+483.xalancbmk+, \verb+473.astar+, \verb+464.h264ref+, and \verb+403.gcc+ also exhibit a clear trend, wherein the slowdown increases as miss rate rises. 
The object counts for these workloads (See Figure~\ref{table:benchmarkgem5data} for details) increase above the largest cache sizes evaluated. However, a high object count alone does not predict performance slowdown, as seen in Figure~\ref{table:benchmarkgem5data}. 
For example, the slowdown for \verb+473.astar+ remains ${\sim}$15-18\% for all cache sizes.
As the miss rates remain constant (${\sim}$19-24\%), the overall performance impact for this workload is modest. Nevertheless, for \verb+483.xalancbmk+, a much smaller miss rate
(${\sim}$1-5\%) 
translates to a much higher slowdown
(${\sim}$8-67\%). 
This is likely due to \verb+483.xalancbmk+'s sensitivity to the misses --- these object loads are critical to application progress. In comparison, \verb+473.astar+ is much less susceptible to its higher miss rate, as the out-of-order processor is able to hide the additional latency of these misses. 
For a 64KiB cache, the hardware \greenfat implementation incurs a 3.92\% overhead as compared to 11.19$\times$ for the software \bluefat implementation.
Although these results are for slightly different benchmarks, and thus a direct one-to-one comparison is not possible.
Nevertheless, it is clear that \greenfat represents a significant performance improvement since major overheads (e.g., DBI) are avoided altogether.
Finally, we remark that our \greenfat implementation represents a relatively conservative hardware extension, in the form of an additional cache and pointer decode step.
Although this is not zero-overhead, our approach is compatible with existing conventional microarchitectures without a radical redesign. This also underscores the opportunity for further optimizations at the hardware level.

Finally, we remark that multi-core architectures could also be supported as future work.
Overall coherence traffic will be low as it is proportional to the allocations and de-allocations and not the number of writes (as with traditional coherence).

\subsection{Limitations}\label{sec:limitations}
\FRP{} protects heap objects under stronger threat models, including brute force attacks (see Sec.~\ref{sec:attacker}).
\FRP{} does not prevent other attacks from outside of this model.
For example, the attacker may acquire a target pointer using a logical error or data race, rather than (\capInvalid).
That said, most exploits depend on a memory error as part of the attack chain, 
and traditional memory errors still represent a significant attack surface.
Like other tools~\cite{nethercote2007valgrind, bruening2011drmem, duck2022redfat}, \bluefat and \greenfat are binary-only, and do not protect non-heap objects, or necessarily objects allocated with {\em Custom Memory Allocators} (CMAs) rather than standard \verb+malloc()+ (depending on the CMA implementation).
However, these are general limitations of binary-only tools (not specific to \FRP{}), and could be lifted if our approach were to be ported to the source level as future work.

\FRP{} is compatible with most off-the-shelf binaries by default.
However, binaries that use architecture-specific tagged pointer representations (that compete with our randomized pointer encoding) may not be compatible.
Examples of incompatibility are rare, and affect {\em all} encoded-pointer defenses (not specific to \FRP{}).

\section{Related Work}
We briefly summarize the related work in this section.
For a more detailed survey on memory error defenses, see~\cite{song19sok}.

\myparagraph{Memory poisoning}
As discussed, memory poisoning is a very common approach implemented by many tools~\cite{purify, asan, seward2005using, hasabnis2012light, bruening2011drmem, nethercote2007valgrind, fuzzan, duck2022redfat, sinha18rest, rezzan}.
The vulnerability to bypass attacks is well-known, and thus, the technique is mainly used for bug detection rather than security hardening.

\myparagraph{Guard pages}
Another pre-existing idea is to insert inaccessible guard pages between memory objects, as used by \emph{efence}~\cite{bruce1993electri}, Archipelago~\cite{lvin2008archipelago}, and GWP-Asan~\cite{gwp-asan}.
Accessing a guard page will trigger a memory fault (\verb+SIGSEGV+).
In addition to consuming large amounts of memory (both virtual and physical), the approach can be vulnerable to bypass in the absence of randomization.

\myparagraph{Use-after-free vulnerabilities}
There exist specialized use-after-free defenses~\cite{kouwe2017dangsan, younan2015free, dang2017oscar,nagarakatte2010cets}.
Oscar~\cite{dang2017oscar} similarly avoids reusing pointers to reallocated memory.
Unlike \FRP{}, Oscar relies on virtual memory tricks, specifically, mapping the same physical page into the virtual address space more than once.
This has overheads, weaker security, and supports fewer reallocations until exhaustion.

\myparagraph{Pointer tagging}
\FRP{} shares some superficial similarities with {\em pointer tagging}.
Here, unused pointer bits are repurposed for metadata, e.g., an object $\mathit{id}$.
For example, HWAsan~\cite{serebryany2018memory} for ARM tags each pointer with a randomized value that is checked on access.
However, pointer tagging essentially encodes the $\mathit{id}$ twice: once explicitly as the tag, and once more implicitly as the address.
\FRP{} encoding supports {\em orders of magnitude} more $\mathit{id}$s (40 versus 16 bits), enabling much stronger security.

\myparagraph{Fat pointers}
Another old idea is to replace pointers with a fat encoding that explicitly stores object metadata,
as used by Safe-C~\cite{safe-c}, CCured~\cite{necula02ccured}, Cyclone~\cite{cyclone}, amongst others.
Like \FRP{}, this approach tracks pointer provenance and avoids pointer confusion.
However, fat pointers violate (\reqCast) and are generally not compatible with existing code without modification.
Softbound~\cite{nagarakatte09softbound} improves over fat pointers by separating metadata, but is not binary compatible and needs recompilation.

\myparagraph{Low-fat pointers}
Another idea is to compress fat pointers into a native address representation~\cite{duck16heap, duck17stack}.
Unlike \FRP{}, this approach can still suffer from false positives in specific cases (e.g., pointer escapes).

\myparagraph{Associative maps}
Tools like CRED~\cite{ruwase2004overflow}, MPX~\cite{intel23manual}, and FuZZan~\cite{fuzzan} store metadata in a tree-based associative map.
However, \FRP{} is the first approach to use an associative map to implement full pointer randomization, with the corresponding security benefits.

\myparagraph{Probabilistic methods}
As discussed, DieHard~\cite{berge06diehard} and Die\-Hard\-er~\cite{gene10dieharder} are the most prominent examples of heap layout randomizers.
However, such randomization is limited to the virtual address space region used by the heap, which is orders of magnitude smaller than the full \FRP{} space.
As such, heap layout randomizers have significantly weaker guarantees compared to \FRP{}.

\myparagraph{Encrypted pointers}
PointGuard~\cite{cowan2003pointguard} {\em encrypts} pointers in memory to mitigate potential overwrites.
Similarly, the \verb+glibc+ \verb+setjmp+ function mangles the \verb+jmp_buf+ structure, and Windows supports an \verb+EncodePointer+ API for encrypting pointers.
However, these schemes are not general-purpose pointer encodings under (\reqAny).
For example, given an encoded $p$, then $\mathtt{DecodePointer}(p{+}1)$ yields a gibberish value. 
Neither the PointGuard, \verb+setjmp+, nor \verb+EncodePointer+ encodings are general-purpose memory error defenses.

{\em Cryptographic Capability Computing} ($C^3$)~\cite{lemay2021c3} is a probabilistic defense system based on {\em partial} pointer encryption but with lower entropy (24 bits, see Table~\ref{tab:sanitizers}).
One advantage of encryption over \FRP{} is that an explicit map ($\mapping$) is not required, since the address can be deterministically derived (i.e., decrypted) from the encrypted pointer.
However, this introduces a significant problem with \emph{use-after-free} (UaF), since a reused address will also reuse the same encrypted pointer value (one-to-one correspondence).
To mitigate this problem, the $C^3$ system includes an 8 bit {\em version} bitfield into the encrypted portion of the address (one-to-256 correspondence).
However, under our stronger threat model, this equates to only \textbf{8 bits} of effective security, significantly weaker than the 24 bits for other classes of memory error. 
By using an explicit mapping, \FRP{} maintains the full (52 bit) security regardless of the error class.

\myparagraph{Pointer provenance and capabilities}
The ambiguity between invalid and valid pointers is closely related to the notion of \emph{pointer provenance}~\cite{memarian2019ptr}.
Some memory error defenses, such as fat pointers~\cite{necula02ccured} and \emph{capability systems} (e.g., CHERI~\cite{cheri} and \textsc{Capstone}~\cite{yu2023capstone}), implement provenance by explicitly tracking per-pointer metadata.
This makes it impossible to {\em forge} an invalid pointer/capability since the metadata for the original object remains associated with the pointer.
However, this tends to break binary compatibility since the per-pointer metadata must be explicitly tracked somehow.
This is further complicated by the lack of type information at the binary-level.

We argue that our fully randomized encoding can be thought of as the first {\em implicit} provenance-preserving pointer representation.
This is a probabilistic and emergent property: two \FRP{} pointer values will have very different bit patterns, and thus it is extremely difficult for the two values to become inadvertently or maliciously confused.
Furthermore, since \FRP{} values are still ordinary machine words, there is no need for explicit object or provenance tracking that would otherwise break binary compatibility.

\section{Conclusion}

Despite significant research into memory error defenses, there is currently no universally adopted, always-enabled solution to prevent these errors in modern systems today. Recent hardware and entropy-based defenses are promising, but the effective security is very low (only 24 bits, at most), which can be bypassed with brute force attacks.

In this paper, we aim to defend against even stronger attack models, including the powerful ability to repeat attack attempts.
Our approach is to design a new pointer encoding that is (1) strongly {\em decoupled} from the underlying memory address, and (2) can be fully randomized---a.k.a. {\em Fully Randomized Pointers} (\FRP{}).
\FRP{} represents pointers as (cryptographically secure) random numbers which are difficult to bypass because pointer values are no longer correlated, nor can they be realistically brute-forced. 

We have implemented \FRP{} in the form of a software prototype (\bluefat{}) as well as a hardware simulation (\greenfat{}).
We show that \bluefat{} provides strong security and is highly compatible with existing binary software, despite using a \textit{radically different} pointer encoding. 
Our \greenfat{} hardware implementation results in a practical performance overhead of 3.92\%.
Preventing memory errors is an unsolved challenge due to the (potentially) conflicting requirements and resulting tradeoffs.
With \FRP{}, we demonstrate that a new tradeoff point---between binary compatibility (without recompilation) and strong bypass resistance---is both practical and feasible. Previous works in hardware capabilities \cite{cheri,yu2023capstone} provide strong bypass resistance but lack compatibility. Existing usable software solutions also weaken security and/or compatibility, which points to \FRP{} as providing strong security, compatibility, and usability.

\section*{Acknowledgments}

This work was funded, in part, by grants from the National Research Foundation (NRF) of Singapore (NRF2018NCR-NCR002),
the Cyber Security Agency (CSA) of Singapore under its National Cybersecurity R\&D Programme (NRF-NCR25-Fuzz-0001),
and the Singapore Ministry of Education (MOE) (T2EP20222-0026).
Any opinions, findings and conclusions, or recommendations expressed in this material are those of the author(s) and do not reflect the views of the NRF or CSA of Singapore.

\bibliographystyle{ACM-Reference-Format}
\bibliography{main}

\end{document}